\documentclass[%
 reprint,
 amsmath,amssymb,
 aps,
 showkeys
]{revtex4-2}

\usepackage{graphicx}% Include figure files
\usepackage{dcolumn}% Align table columns on decimal point
\usepackage{bm}% bold math
\usepackage[T1]{fontenc}
\usepackage{xcolor}

\DeclareUnicodeCharacter{0229}{ę}
\DeclareUnicodeCharacter{0229}{\k{e}}
\DeclareUnicodeCharacter{202F}{\textendash}

\newcommand{\beginsupplement}{%
        \setcounter{table}{0}
        \renewcommand{\thetable}{S\arabic{table}}%
        \setcounter{figure}{0}
        \renewcommand{\thefigure}{S\arabic{figure}}%
     }

\begin{document}

\preprint{APS/123-QED}

\title{Temperature induced giant shift of phonon energy in epitaxial boron nitride layers}

\author{Jakub Iwański}
\email{Jakub.Iwanski@fuw.edu.pl}
\author{Piotr Tatarczak}
\author{Mateusz Tokarczyk}
\author{Grzegorz Kowalski}
\author{\\Aleksandra K. Dąbrowska}
\author{Johannes Binder}
\author{Roman St\k{e}pniewski}
\author{Andrzej Wysmołek}
\affiliation{Faculty of Physics, University of Warsaw, ul. Pasteura 5, 02-093 Warsaw, Poland}
%\date{\today}

\begin{abstract}

The recent progress in the growth of large-area boron nitride epilayers opens up new possibilities for future applications. However, it remains largely unclear how weakly attached two-dimensional BN layers interact with their substrate and how their properties are influenced by defects. In this work, we investigate hBN layers grown by Metal Organic Vapor Phase Epitaxy (MOVPE) using Fourier-transform Infrared (FTIR) spectroscopy in the temperature range of 160-540~K. Our measurements reveal strong differences in the character of layer-substrate interaction for as-grown and delaminated epitaxial layers. A much weaker interaction of as-grown layers is explained by wrinkles formation that reduces strain at the layer-substrate interface, which for layers transferred to other substrates occurs only in a limited temperature range. The most striking result is the observation of a giant increase in the $E_{1u}$ phonon energy of up to $\sim6$~cm$^{-1}$ in a narrow temperature range. We show that the amplitude and temperature range of the anomaly is strongly modified by UV light illumination. The observed giant effect is explained in terms of strain generation resulting from charge redistribution between shallow traps and different defects, which can be interpreted as a result of strong electron-phonon coupling in hBN. The observed narrow temperature range of the anomaly indicates that the effect may be further enhanced for example by electrostrictive effects, expected for sp$^2$ boron nitride.

\end{abstract}

\keywords{epitaxial boron nitride, hBN, FTIR, MOVPE, $E_{1u}$ phonon, defect charged states}

\maketitle

\section{INTRODUCTION}

	Hexagonal boron nitride (hBN) is a wide bandgap layered material \cite{Zunger1976} with strong sp$^2$ covalent bonds in-plane and weak van der Waals bonds between adjacent layers \cite{Ohba2001}. An additional asset of hBN is the high thermal and chemical stability \cite{Viana2020, Chen2004, Kostoglou2015, Sevik2011}. These properties make hBN an important component in many novel van der Waals heterostructures \cite{Jadczak2021, Prazdnichnykh2021, Li2017, Binder2017} or as an encapsulating layer that increases the durability of other less durable materials such as perovskites \cite{Seitz2019} or transition metal dichalcogenides \cite{Lee2015, Ahn2016}.
	
	However, the effective usage of many systems and microdevices in everyday life is dependent on the possibility to be scaled up. The solution for this problem are epitaxial techniques which enable few-inch area layer growth. In case of hBN the structural and optical quality of the epitaxial layers is still inferior to those of better-studied flakes exfoliated from the bulk crystal. The difference between epitaxial and bulk hBN is also observed in the interaction with other materials \cite{Ludwiczak2021, Pacuski2020}. Hence, in order to combine epitaxial hBN layers with other materials it is important to understand the different properties of defects in epitaxial hBN and the mechanisms of their creation as well as the rules that govern interactions between epitaxial hBN and other materials.

	In this work, we present $E_{1u}$ hBN phonon energy-temperature dependences measured with Fourier-transform infrared (FTIR) spectroscopy. We measured delaminated and as-grown epitaxial layers. For as-grown samples characteristic wrinkles are created during cooling after growth, because of the relaxation of the accumulated strain energy caused by the negative thermal expansion coefficient of hBN \cite{Yates1975}. We revealed a significant reduction in the layer-substrate interaction, which is a consequence of wrinkles creation. In addition, we observe an anomalous, giant upshift of phonon energy that occurs in a narrow temperature range. This effect is explained in terms of charge redistribution of hBN defects. The observed effects highlight that defects in hBN play not only an important role for electrical conductivity, mid-gap photoluminescence or single photon emission, but can also strongly affect the phonon modes in a complex manner. 

\section{METHODS}
\subsection{Samples}
The boron nitride samples used in this work were obtained with Metal Organic Vapor Phase Epitaxy (MOVPE) using an Aixtron CCS 3$\times$2” system. Two-inch sapphire wafers were used as substrates, and ammonia and triethylboron (TEB) were used as precursors of nitrogen and boron, respectively. In most processes, hydrogen was used as a carrier gas, but a few samples were obtained using a nitrogen atmosphere. Thin epitaxial boron nitride samples (from atomic monolayer to 70~nm), oriented parallel to the sapphire surface, were obtained by three growth regimes: Continuous Flow Growth \cite{arxiv,Chugh2018}, Flow Modulation Epitaxy \cite{Chugh2018, Kobayashi2008} and Two-stage Epitaxy \cite{Dabrowska2020}. Due to a pre-growth nitridation step a thin AlN layer is formed at the sapphire/hBN interface. For measurements, we prepared $\sim8\times 8$~mm pieces of investigated hBN layers. We are aware that our samples may consist of boron and nitrogen atoms bonded with sp$^2$ bonding in different phases such as hexagonal (hBN), rhombohedral (rBN), turbostratic (tBN), Bernal stacking (bBN) \cite{Dabrowska2020, Moret2021}. However, to simplify the notation in the present work, we will use hBN having in mind all layered boron nitride phases.

For temperature-dependent FTIR measurements, we chose a 41~nm thick hBN layer grown in Flow Modulation Epitaxy regime and a 20~nm thick hBN layer grown in Two-stage Epitaxy regime. The thicker layer was also delaminated and deposited onto a new sapphire substrate to obtain information about the layer-substrate interaction. The delamination procedure is already described in the literature \cite{Iwanski2021, Chugh2018}.

\subsection{Experimental details}

	X-ray measurements were performed using a Panalytical X’pert diffractometer equipped with a Cu sealed X-ray tube and a parallel beam Bragg X-ray mirror. For all samples, X-ray reflectometry (XRR) measurements were taken ($\omega/2\theta$ scans). %Because the measurements are taken at small incidence angles, the signal obtained comes from a large area of the sample examined (information is averaged from at over a dozen mm2). 
	To obtain the thickness of the hBN layer from the XRR measurements, the PANalytical X’Pert Reflectivity software package was used. This program makes use of the recursive Parratt formalism for reflectivity \cite{Parratt1954, fewster2003, Fewster1996}. In the fitting procedure, the thickness, density, and roughness of the hBN and AlN layers were taken into account, as well as the roughness of the sapphire substrate. It should be pointed out that in the calculations only a one-dimensional model is considered.

	A Thermo Scientific Nicolet Continu$\mu$m Infrared Microscope with a 32x Schwarzschild infinity corrected objective (NA 0.65) was used for FTIR reflection measurements. All samples presented were measured at 3 points approximately 3~mm apart with an illuminated area of 70$\times$70 $\mu$m. The spectra were collected in the range of 650 to 4000~cm$^{-1}$ with a resolution of 0.964~cm$^{-1}$ for samples thinner than 4~nm and 0.482~cm$^{-1}$ for the rest of the samples. The light beam was perpendicular to the surface of the sample.

	A Linkam THMS350EV cryostat was used for temperature dependence measurements. The real temperature value was measured with a thermocouple soldered to the sample surface with indium. The calibration allowed for a range of possible temperatures between 157-538~K. Before each temperature-dependent measurement, the sample was measured at room temperature on 6 different locations to be sure that the sample is isotropic.
	
	For UV sample irradiation, we used an EQ-99X LDLS lamp emitting light in the range 170-2500~nm.

\subsection{Characterization method}

To characterize hBN layers, we simulate the whole FTIR spectrum. To do so, we assume that the material is a medium composed of harmonic damped oscillators. This allows us to introduce Dynamic Dielectric Function (DDF). In the function there are parameters such as oscillator self-energy $\omega_0$ and its damping parameter $\gamma$ which provide us with information about the  phonon properties in the material. In our analysis method, we consider a thin layer on a thick substrate, also described by its DDF. This approach also allows to describe heterostrucutres that consist of several layers or materials. It is possible because of the phase difference analysis that comes from the interference of light beams reflecting on the subsequent interfaces. This method allows to obtain information about the thickness of the layer $d$ and interface roughness $\sigma$ considered for $d>100$~nm. All the details about the characterization method are available in the Supplementary Materials.

\begin{figure}[t]
\includegraphics[width=\columnwidth]{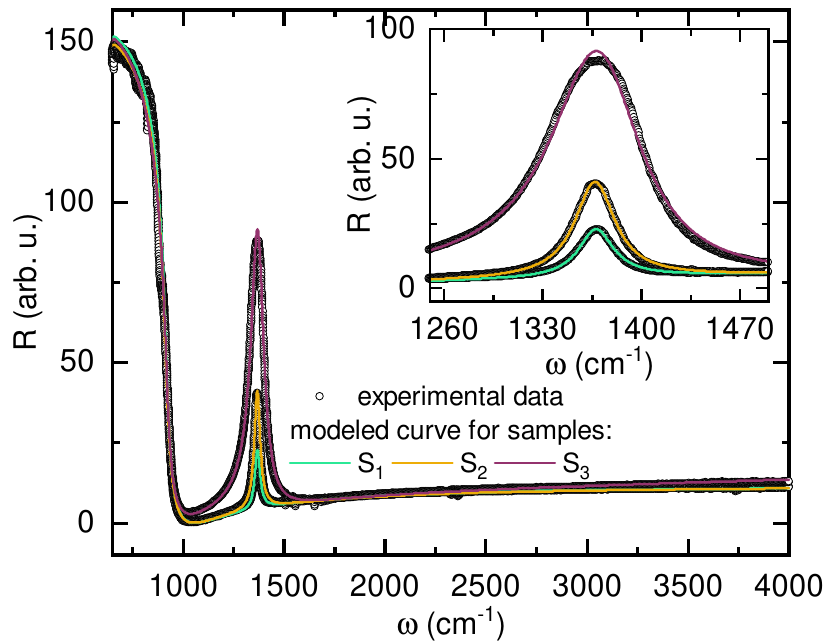}
\caption{\label{fig:ex_fit} FTIR spectra measured at room temperature for three different samples ($S_1, S_2, S_3$) of epitaxial hBN layers on a Al$_2$O$_3$ substrate (circles) with modeled curves (lines). The inset shows zoom of the hBN $E_{1u}$ vibrational mode peak.}
\end{figure}

\begin{table*}[t]
\caption{\label{tab:par_fit}
Best fit parameters of curves modeled for FTIR spectra in FIG.~\ref{fig:ex_fit}}
\begin{ruledtabular}
\begin{tabular}{lccccc}
Sample&$\omega_{BN}$ (cm$^{-1}$)&$\gamma_{BN}$ (cm$^{-1}$)&$d_{BN}$ (nm)&$\omega_{sapph}$ (cm$^{-1}$)&$\gamma_{sapph}$ (cm$^{-1}$)\\
\hline
$S_1$ & 1367.69 & 22.60 & 8.47 & 917.45 & 37.18 \\
$S_2$ & 1367.15 & 20.25 & 15.66 & 917.58 & 38.75 \\
$S_3$ & 1368.66 & 22.40 & 60.28 & 918.69 & 33.72 \\
\end{tabular}
\end{ruledtabular}
\end{table*}

In FIG.~\ref{fig:ex_fit} we present an exemplary spectrum modelling for three different samples. The samples were grown using three different growth conditions that resulted in different material properties (i.e. strain, amount of defects, thickness). The modelling works well for each of the samples. The values of the modelled parameters are presented in TAB.~\ref{tab:par_fit}. The peak position at about 1367~cm$^{-1}$ corresponds to the $E_{1u}$ phonon mode of sp$^2$ hybridized boron nitride \cite{Geick1966}. The high quality is  manifested by the values of the $\gamma_{BN}$ parameter for the  $E_{1u}$ mode, which are relatively small compared to the values reported for epitaxial layers \cite{Chugh2018, Caban2020, Bera2020, WANG2021677}. The inset in FIG.~\ref{fig:ex_fit} shows that the $E_{1u}$ peak intensity directly depends on the thickness of hBN. The high reflectivity below 1000~cm$^{-1}$ is connected to the presence of sapphire substrate \cite{Lee2014}. One can see that the modeled curves in FIG.~\ref{fig:ex_fit} describe experimental data almost entirely. All important features of the spectrum are reproduced and the values in TAB.~\ref{tab:par_fit} are very precisely determined with relative error values $\simeq 0,01\%$ for $\omega_{BN}$ and $\simeq 1\%$ for $\gamma_{BN}$ and $d_{BN}$. This precision enables to detect smallest deviations in strain, quality and thickness of epitaxial layers. Furthermore, due to modeling parameters related to sapphire ($\omega_{sapph}$, $\gamma_{sapph}$), one can monitor the quality of the substrate on which hBN was grown or deposited.

The principal advantage of the described method over the very basic peak Lorentzian fitting is the access to the physical properties of the sample, even if the sample structure is complex. The absorption shape modified by the damping parameter $\gamma$ which influences the peak position or layer thickness is out of range for simple fitting. The thickness results coming from our method are in very good agreement with the results obtained with XRR for the samples in the range 2-70~nm (see Supplementary Materials FIG.~S1). The method can be applied for samples in the thickness range from a single atomic layer to several $\mu$m (see Supplementary Materials FIG.~S2).

\section{RESULTS AND DISCUSSION}

\begin{figure*}[htbp]
\includegraphics[width=2\columnwidth]{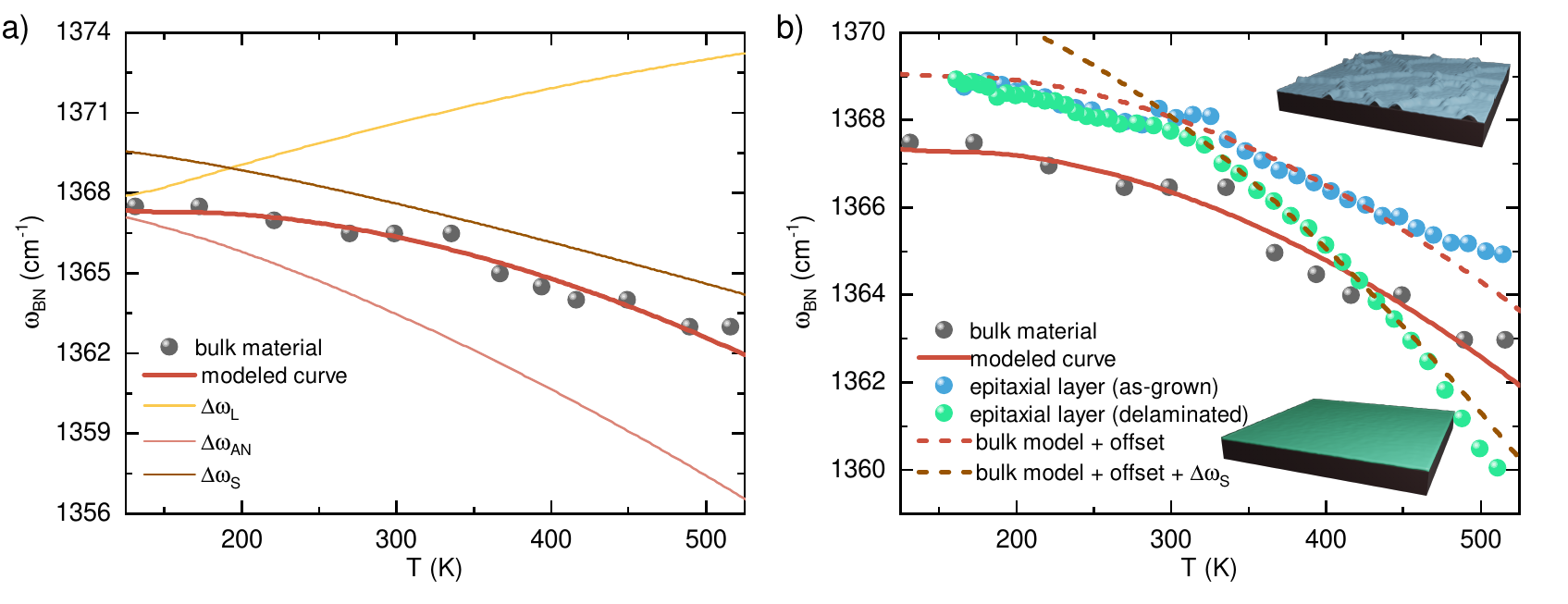}
\caption{\label{fig:model_temp} hBN phonon energy as a function of temperature. Grey spheres represent data for bulk material obtained by Segura et al. \cite{Segura2020}. a) thinner lines represent contribution to the general trend that comes from lattice thermal expansion ($\Delta\omega_L$) (yellow), anharmonic phonon processes ($\Delta\omega_{AN}$) (pink) and material-substrate interaction ($\Delta\omega_S$) (brown) considered for non-bulk samples. The thicker dark red line is described by the introduced model and does not include material-substrate interaction. b) blue and green spheres represent data for as-grown and delaminated layer respectively. The dashed dark red line is created by adding a constant offset to the modeled curve for bulk material. The dashed brown line is created by adding offset and $\Delta\omega_S$ contribution to the modeled curve for bulk material. The pictures in the insets schematically show the appearance of as-grown (wrinkled) and delaminated (flat) layers.}
\end{figure*}

Since the phonon energy is a good indicator for strain \cite{Bera2020}, we investigated as-grown and delaminated samples. Due to the delamination process, both samples should exhibit different layer-substrate interactions. The interactions change with temperature because of the difference in the thermal expansion coefficients of the layer and the substrate. As we provided a reliable method for the identification of the phonon energy of hBN, it is possible to analyze the temperature dependencies of the lattice vibrations of the epitaxial layers $\omega_{BN}(T)$. As we present in FIG.~\ref{fig:model_temp}, the phonon energy for as-grown and delaminated layers follows different trends. To be able to extract information about layer-substrate interactions, we need to know what influences this $\omega_{BN}(T)$ dependency. It is governed by 4 dominant factors: the undisturbed phonon energy at 0~K $\omega_0$, the lattice constant change (thermal expansion) $\Delta \omega_L$, the anharmonic phonon decay $\Delta\omega_{AN}$, the layer-substrate interaction $\Delta\omega_S$ caused by the difference in thermal expansion coefficients of layer and substrate:
\begin{equation}
    \omega_{BN}(T) = \omega_0 + \Delta\omega_L(T) + \Delta\omega_{AN}(T) + \Delta\omega_S(T) \label{eq:omegaBN}    
\end{equation}
Taking the hBN thermal expansion coefficient provided by Yates et al. \cite{Yates1975}, and using it for the Density Functional Theory (DFT) calculations as presented by Cusco et al. \cite{Cusco2016}, one can obtain the energy change due to the thermal expansion coefficient:
\begin{equation}
    \Delta\omega_L(T) = -2.347+2.137\times 10^{-2} T - 1.177\times 10^{-5} T^2
\end{equation}
According to DFT calculations, the $E_{1u}$ phonon has two dominant decay paths: three-phonon process and four-phonon process \cite{Segura2020}. Both give the following contribution to $\Delta\omega_{AN}$ \cite{Balkanski1983}:
\begin{widetext}
\begin{eqnarray}
    \Delta\omega_{AN}(T) &=& A\left( 1+ \frac{1}{e^{\theta_1/T}-1} + \frac{1}{e^{(\theta_0-\theta_1)/T}-1} \right) + \nonumber\\
    &+& B\left(1+\frac{2}{e^{\theta_1/T} - 1} + \frac{1}{e^{(\theta_0-2\theta_1)/T}-1} +\frac{2}{\left(e^{\theta_1/T} - 1\right) (e^{(\theta_0-2\theta_1)/T)}-1)} +\frac{2}{(e^{\theta_1/T} - 1)^2} \right) \label{eq:omega_an}
\end{eqnarray}
\end{widetext}
where $A, B$ - constants are related to three- and four-phonon processes respectively, $\theta_i = (hc/k)\omega_i$ ($h$ - Planck constant, $c$ - speed of light, $k$ - Boltzman constant, $\omega_i$ - phonon energy). In Eq.~\ref{eq:omega_an} we assumed that in both three- and four-phonon processes, there is the same energy decay product. This assumption is justified since in both decays the produced lower energy phonons are along the $M$-$K$ line in the dispersion relation, which is flat. The change related to the layer interaction with the substrate can be described with the following formula:
\begin{equation}
    \Delta\omega_S(T) = \beta_{BN}\int_{T_{r}}^{T} \alpha_{sapphire}(T') - \alpha_{BN}(T') dT'
\end{equation}
where $\alpha_{sapphire,BN}$ is the thermal expansion coefficient for sapphire and BN, $\beta_{BN}$ is the hBN Gr{\"u}neisen parameter and $T_r$ is the temperature at which the layer on the substrate is relaxed. For the as-grown layer, it is the growth temperature, and for the delaminated layer it is room temperature. All the described contributions of $\omega_{BN}(T)$ dependency are plotted in FIG.~\ref{fig:model_temp}a.

Using the model that does not include the layer-substrate interaction, we can fit the line that follows experimental data for bulk material reported by Segura et al. \cite{Segura2020}. This gives the following values of parameters: $\omega_0 = 1376$~cm$^{-1}$, $A = -7.25$~cm$^{-1}$, $B = -1.19$~cm$^{-1}$ and $\theta_1 = 369.78$~K. It corresponds to the decay to two phonons of energies 257~cm$^{-1}$ and 1119~cm$^{-1}$ in the three-phonon process, two phonons of energy 257~cm$^{-1}$ and one phonon of energy 862~cm$^{-1}$ in four-phonon process. These results are in good agreement with the literature values \cite{Geick1966, Cusco2016}.

In FIG.~\ref{fig:model_temp}b we present the dependence of $\omega_{BN}(T)$ for the epitaxial layers. Both sets of results were collected for samples grown in the same MOVPE process. However, one of them was delaminated and transferred to a new sapphire substrate. The method used for delamination and transfer is as described in the literature \cite{Iwanski2021}. The blue spheres related to the as-grown sample follow the same trend as the data for bulk material, which illustrates a dark red dashed line. The line is drawn by adding a constant offset to the modeled curve for bulk hBN. The shift towards higher energies is due to a higher amount of defects in the epitaxial layer, which leads to additional strain. This additional effect does not allow for reliable parameters modeling because of an unknown temperature-dependent behaviour of defects. It is suggested that the defect contribution should be linear \cite{Bera2020}. Adding a further linear or quadratic term to the model would strongly influence other parameters which at this point are in good agreement with the literature values. We later show that defects indeed play an important role and that the assumption of a linear defect contribution is not always justified. For this reason, we provide only qualitative results.

The green spheres related to the delaminated sample exhibit a more significant decrease in phonon energy with increasing temperature. This decrease is most likely caused by a stronger layer-substrate interaction, as suggested by the brown dashed line which was created by adding $\Delta\omega_S$ term to the dark red dashed line. To cover the experimental data in a as wide as possible range, we assumed the hBN Gr{\"u}neisen parameter $\beta_{BN} = -21$~cm$^{-1}$/\%. This value is over two times lower than reported for hBN before \cite{Androulidakis2018}. The main reason for the change of layer-substrate interaction $\Delta\omega_S$ before and after delamination is the presence and lack of characteristic wrinkles. They are created during the cooling of the material after the growth process and can be removed in the delamination process \cite{Iwanski2021}. The results presented in FIG.~\ref{fig:model_temp}b imply that in the as-grown layer, the wrinkles can compensate for the layer-substrate strain. Consequently, the blue spheres run parallel to the modeled curve for the bulk material, as the dark red dashed line suggests. The strain in the case of the as-grown epitaxial layer is eliminated by excess material stored in the wrinkles. The same behaviour is observed for the delaminated layer below room temperature at which the delamination process occurred. When cooling the relaxed delaminated layer, the wrinkles can be created and compensate for the layer-substrate strain. Consequently, the $\omega_{BN}(T)$ trend below room temperature is similar for both delaminated and as-grown layers. Equally, the effect of the decrease in phonon energy related to  $\Delta\omega_S$ contribution for as-grown layers could be observed at much higher temperatures, which are closer to the growth temperature at which the wrinkles should disappear.

\begin{figure}[htbp]
{\includegraphics{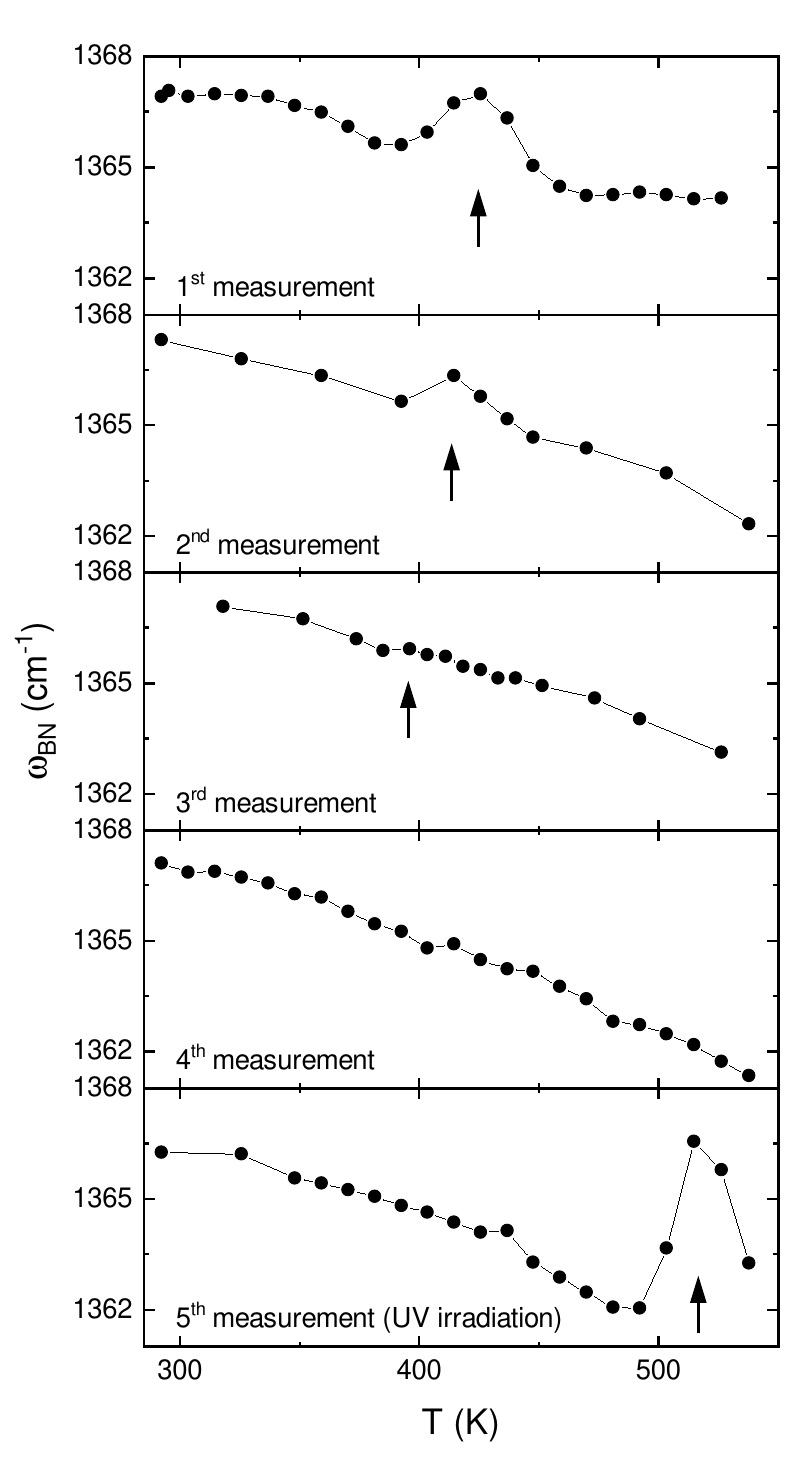}}
 \caption{\label{fig:anomaly} $E_{1u}$ phonon energy as a function of temperature for the 20~nm thick sample grown in the Two-stage Epitaxy regime.  Arrows indicate the anomalous behaviour. The initial four temperature-dependent measurements lead to a softening of the anomalous phonon energy shift. After UV irradiation the anomalous effect is enhanced. Lines are guide to the eye only.}
\end{figure}

The most striking result that emerges from our measurements is a giant, anomalous shift of the $E_{1u}$ phonon energy which we observed for some samples. In FIG.~\ref{fig:anomaly} we present the $\omega_{BN}(T)$ dependence for a 20~nm thick layer grown in the Two-stage Epitaxy regime. One can notice an anomalous increase of phonon energy at 425~K for the 1$^{\mathrm{st}}$ measurement. This increase cannot be explained by the presented model. When performing consecutive temperature-dependent measurements, the anomaly softens and its maximum slightly moves towards lower temperatures as presented for 2$^{\mathrm{nd}}$ and 3$^{\mathrm{rd}}$ measurement. In the 4$^{\mathrm{th}}$ temperature run, no anomalous phonon behaviour is observed. To make sure that the effect is not an artifact of our data analysis method, we performed basic peak Lorentzian fitting on the measured data set. In this approach, the anomaly is still observed at the same temperatures (see Supplementary Materials).

We also observed the described anomalous behavior of $\omega_{BN}(T)$ dependence for other samples grown under different regimes under NH$_3$-rich conditions before and after the delamination process (see Supplementary Materials). The anomaly occurrence for different samples implies that the phonon upshift is not an accidental effect related to one specific sample, but can be reproduced when a special set of growth parameters is applied. The observation of the effect occurrence for both as-grown and delaminated layers proves that it does not come from the substrate or thin AlN layer on Al$_2$O$_3$-hBN interface but it is attributed to the hBN layer itself.

In the following measurements presented in FIG.~\ref{fig:anomaly}, there is a more steep decrease in energy with increasing temperature. The same effect is observed when we compare the results for as-grown and delaminated layers in FIG.~\ref{fig:model_temp}b. In this case, it is most likely related to changes in the layer wrinkle pattern. As the result suggests, the wrinkle pattern can be changed for temperatures much lower than the growth temperature. Wrinkles rearrangement leads to an increase in the layer-substrate interaction contribution what is observed in the energy downshift for higher temperatures, as suggested by the brown curve in FIG.~\ref{fig:model_temp}a. 

\begin{figure}[bp]
{\includegraphics{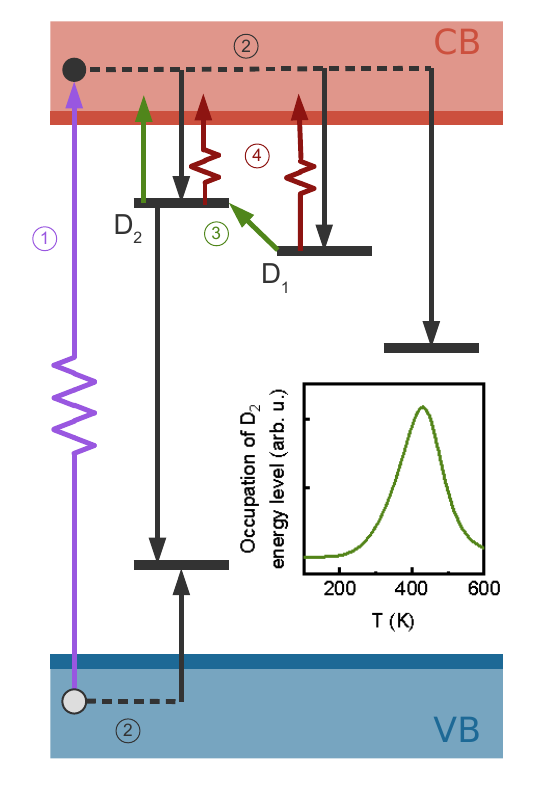}}
 \caption{\label{fig:model} Schematic drawing of the most probable carriers transitions that explain the anomalous phonon behaviour (the distances between defect levels and bands are not in scale). The inset shows a simulation (see Supplementary Materials) of occupation of the middle energy level $D_2$.}
\end{figure}

Once the anomaly vanishes after consecutive temperature runs, we can restore it by UV irradiation. The anomaly measured after UV irradiation in the 5$^{\mathrm{th}}$ run is much stronger and appears at higher temperature as presented in FIG.~\ref{fig:anomaly}. The increase in phonon energy is $\sim6$~cm$^{-1}$. We propose the mechanism presented in FIG.~\ref{fig:model} to explain the observed phenomenon. When the sample is exposed to UV light, electrons from the valence band are transferred to the conduction band (CB) (\textcircled{\raisebox{-0.9pt}{1}}). Then they can be trapped on a shallow defect $D_1$ just below the conduction band (\textcircled{\raisebox{-0.9pt}{2}}). Here, we assume that $D_1$ defect energy level is less than 1~eV below the conduction band, which has high degeneracy. The $D_1$ state can be filled during the growth process as well. These kinds of traps can store carriers for a significant time \cite{Wlasny2019}. If the material has an additional defect state $D_2$ in between state $D_1$ and the CB, we can change the occupation probability of all three states during the measurement by changing the temperature (\textcircled{\raisebox{-0.9pt}{3}}). For a specific temperature, the occupation of the $D_2$ defect state reaches the maximum value. The change in electrical state of $D_2$ leads to the expansion of the defect. Consequently, an inhomogeneous compressive strain is introduced. For such conditions, the phonon energy increases. This dependence between the electric state of the defect and the phonon energy suggests the presence of a strong electron-phonon coupling in those epitaxial hBN, in particular as comparing to the bulk material \cite{Wlasny2019}. A further temperature increase transfers electrons to the conduction band. Because the traps are situated close to the conduction band, they can be refilled with electrons. Then the anomalous effect is repeated for the next temperature-dependent measurement until we empty most of the shallow traps, and electrons occupy deeper donor levels. Further evidence for the presented mechanism is provided by the fact that the anomaly is observed in both directions, for an increase as well as for a consecutive decrease in temperature (see Supplementary Materials). We do not exclude that the presented mechanism could be occurring symmetrically for the valence band and acceptor states.

The defects properties in hBN are complicated and not fully understood. For this reason, it is hard to pinpoint defects $D_1$ and $D_2$. However, we have candidates that meet the described criteria. It has been reported that the energy levels of $O_N$ are shallow traps placed below the conduction band in hBN \cite{Weston2018, Katzir1975, Vokhmintsev2017, Vokhmintsev2019, Vokhmintsev2021}. At very high temperature during the growth process, oxygen atoms diffuse and desorp from the Al$_2$O$_3$ substrate \cite{Matsuoka2018}. Consequently, the substrate can be a source of oxygen atoms in the hBN layer that replace nitrogen atoms. Such a defect has a low formation energy \cite{Weston2018}, so it should be present in all high-temperature grown epitaxial layers. The $O_N$ defect is therefore a good candidate for the $D_1$ defect.

The studied hBN layer was grown in the Two-stage Epitaxy regime in NH$_3$ rich conditions \cite{Dabrowska2020} which supports the creation of $V_B$ \cite{Weston2018}. $V_B$ in 1- and 2- charge states increases the distortion between nitrogen atoms \cite{Weston2018} which leads to lattice compressive strain in vicinity of the defect. Moreover, $V_B$ has an energy similar to that of the $O_N$ defect and the $V_B-O_N$ complex \cite{Weston2018}. Therefore, we speculate that the $D_2$ defect can be a negatively charged boron vacancy.

The anomaly presented in FIG.~\ref{fig:anomaly} occurs in a narrower range of temperatures in comparison to the simulation in the inset in FIG.~\ref{fig:model}. That suggests that there are other effects that enhance the anomalous phonon energy behaviour. Such an effect could be the piezoelectric response of boron nitride which is observed for different phases of sp$^2$-BN \cite{Kundalwal2021, Michel2011, Brazhe2020}. Such non-centrosymetric phases, which exhibit piezoelectricity, are present in epitaxial boron nitride layers \cite{Moret2021}. Moreover, there is evidence of a strong electrostrictive effect in boron nitride nanotubes \cite{Kang2015, Dai2009}. Electrostriction depends quadratically on the electric field. Therefore, the reverse of the field does not reverse the direction of strain. Consequently, the strain does not disappear when considering diferent randomly arranged BN grains. The deformation related to the electrostrictive effect is much greater compared to traditional piezoceramics \cite{Dai2009}. Consequently, the electric field induced by charge redistribution on defect states would significantly influence the phonon energy change as observed in the experiment.

\section{CONCLUSIONS}

In this work, we studied the thermal response of the $E_{1u}$ phonon mode in boron nitride epitaxial layers using Fourier-transform infrared spectroscopy. The developed FTIR spectra analysis provided reliable information about the physical properties of epitaxial layers which are in good agreement with the X-ray reflectometry results. Both methods are complementary. However, our method allows to analyze a wide range of samples, from a single atomic monolayer to a few tens of $\mu$m.  The upper thickness limit depends on the quality of the sample. We can also obtain information about surface roughness or material porosity. The method can be easily extended to other materials and structures.

In our study, we investigated as-grown as well as delaminated layers. The results of temperature-dependent measurements emphasize the importance of layer wrinkles in the layer-substrate interaction. The wrinkles significantly reduce the strain on the interface. Consequently, the behaviour of the epitaxial layer is similar to that of the bulk material. 

Furthermore, we observed strong electron-phonon coupling, which can lead to a phonon energy increase up to 6~cm$^{-1}$. This anomalous behaviour vanishes after sample heating and cooling cycles. Importantly, the process can be enhanced by UV irradiation. This effect can be explained by the presence of defects ($D_1$, $D_2$) close to the conduction band. The exact nature of such defects is still unknown. However, we speculate that these defects could be $O_N$ and $V_B$ or their complexes. By increasing the occupation of the upper $D_2$ energy state, we introduce an inhomogeneous compressive strain which leads to an anomalous increase in phonon energy. Most likely, the observed anomaly is enhanced by other effects such as piezoelectricity and electrostriction. However, more research is needed to investigate the exact properties of shallow impurity states and their influence on the properties of hBN.

\begin{acknowledgments}
This work was supported by the National Science Centre, Poland, under the decisions 2019/33/B/ST5/02766 and 2020/39/D/ST7/02811.
\end{acknowledgments}

\bibliography{bibliography}%

\newpage
\beginsupplement
\begin{widetext}
\section*{Supplementary materials}

\section*{FTIR spectra analysis method}

To simulate the whole FTIR spectrum, we assume that the medium is composed of harmonic damped oscillators. Then Dynamic Dielectric Function (DDF) for hBN $\varepsilon_{BN}$ is given by \cite{Geick1966}:
\begin{equation}
    \varepsilon_{BN}(\omega) = \varepsilon_{\infty} + \frac{s_{1}^2}{\omega_{1}^2 - \omega^2 - i\gamma_{1}\omega} + \frac{s_{2}^2}{\omega_{2}^2 - \omega^2 - i\gamma_{2}\omega} \label{eq:eps_BN}
\end{equation}
where $\varepsilon_{\infty}= 4.95$, $s_1^2 = 1.23\times10^5$~cm$^{-2}$, $s_2^2 = 3.49\times10^5$~cm$^{-2}$, $\omega_1$ and $\omega_2$ are the frequencies of the normal modes $A_{2u}$ and $E_{1u}$ respectively and $\gamma_{1,2}$ are the corresponding damping parameters. For sapphire dielectric function $\varepsilon_{sapph}$ is described as follows \cite{Lee2014}:
\begin{equation}
    \varepsilon_{sapph}(\omega) = \varepsilon_{\infty, sapph}\prod_{k=1}^4 \frac{\omega_{LO,k}^2 - \omega^2 - i\omega\gamma_{LO,k}}{\omega_{TO,k}^2 - \omega^2 - i\omega\gamma_{TO,k}} 
\end{equation}
where  $\omega_{LO,k}$ and $\omega_{TO,k}$ are the frequencies of longitudinal and transverse optic normal modes in sapphire,  $\gamma_{LO,k}$ and $\gamma_{TO,k}$ are their damping parameters. $\varepsilon_{\infty, sapph}$ is given by the formula \cite{Malitson1962}:
\begin{equation}
    \varepsilon_{\infty,sapph}(\omega) = 1 + \frac{A_1\omega_{1,sapph}^2}{\omega_{1,sapph}^2 - \omega^2} + \frac{A_2\omega_{2,sapph}^2}{\omega_{2,sapph}^2 - \omega^2}
\end{equation}
The values of the parameters connected to sapphire are taken from the papers by Lee et al. \cite{Lee2014} and Malitson \cite{Malitson1962}. In the next step, we assume that incident light is reflected from air-hBN and hBN-Al$_2$O$_3$ interfaces and the light transmitted into the sapphire is fully absorbed and does not interact with the beam reflected on the interfaces. The light beams reflected from the following interfaces interfere with each other. That leads to the formula for the reflection coefficient $r$ for the structure of a thin layer grown on the sapphire substrate \cite{Knittl1976}.
\begin{gather}
 r = \frac{r_1 + r_2 e^{i\Delta\varphi}}{1+r_1r_2e^{i\Delta\varphi}} \\
 r_1 = \frac{1-\widetilde{n}_{BN}}{1+\widetilde{n}_{BN}}, \qquad r_2 = \frac{\widetilde{n}_{BN} - \widetilde{n}_{Al_2O_3}}{\widetilde{n}_{BN} + \widetilde{n}_{Al_2O_3}}
\end{gather}
where $\Delta\varphi = 4\pi dn_{BN}$ is the optical path difference for light that penetrates the hBN layer, $d$ - thickness of the hBN layer and $\widetilde{n}_{BN}$, $\widetilde{n}_{Al_2O_3}$ are complex refractive indices of hBN and sapphire described by the square root of their dielectric functions ($\widetilde{n_i} = \sqrt{\varepsilon_i(\omega)}$). Finally, by calculating the square of the absolute value of the reflection coefficient, it is possible to simulate the reflectivity curve of the analyzed structure ($R=\alpha|r|^2$, where $\alpha$ is a normalization factor).

To verify the propriety of the results coming from the described FTIR spectra analysis method, we compared them with the results obtained with an independent measurement technique, which is X-ray reflectometry. The values of hBN thickness $d$ for the samples in range $2-70$~nm obtained from FTIR ($d_{FTIR}$) and XRR ($d_{XRR}$) measurements are plotted in FIG.~\ref{fig:d_FTIRvsd_XRR}. The $d_{FTIR}$ values presented are the average of values coming from the 3 measured areas. The error bars are their standard deviation. The uncertainty of the $d_{XRR}$ fitted values is about 1~nm. The points lie on the dashed line described by the formula $d_{FTIR}=d_{XRR}$. This indicates a very good agreement between both methods over a wide range of sample thickness.

\begin{figure}[hbtp]
\includegraphics[width=0.5\textwidth]{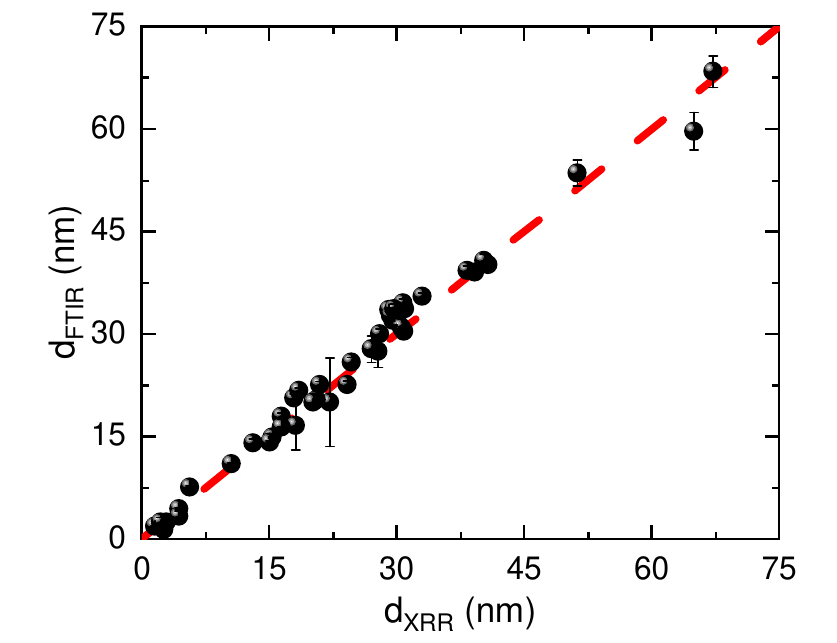}
\caption{\label{fig:d_FTIRvsd_XRR}hBN thickness values obtained with FTIR spectra analysis method as a function of hBN thickness values obtained with X-ray reflectometry measurement (black spheres) and $d_{FTIR} = d_{XRR}$ dashed red line.}
\end{figure}

The next step in the introduction of the method is to define its limitations. In FIG.~\ref{fig:FTIR_limits}a we present the spectrum for the thinnest measured sample. The peak related to the $E_{1u}$ mode in hBN is barely visible. Implementation of our method revealed $d_{BN} = 0.25$~nm. This value is slightly lower than the thickness of the hBN monolayer \cite{Chubarov2018}. This is because the measured area of $70\times 70$~$\mu$m is not fully covered by the hBN layer. However, the data presented in FIG.~\ref{fig:FTIR_limits}a proves that our method has no lower limitations in the context of sp$^2$-BN detection and enables atomic monolayer characterization.

The spectrum for the thickest measured layer is presented in FIG.~\ref{fig:FTIR_limits}b. For this sample, we measured spectra in the range up to 6500~cm$^{-1}$ for better observation of characteristic oscillations related to interference within the hBN layer. Due to the roughness and porosity of thick epitaxial layers, we cannot use our model to describe the experimental data. However, we can introduce some amendments to the model. To take into account the roughness $\sigma$, we change the expression for the reflection coefficient as follows:
\begin{equation}
     r = \frac{r_1e^{-8(\pi\omega\sigma)^2} + r_2Ae^{i4\pi\omega n_{BN}d}}{1 + r_1r_2e^{-8(\pi\omega\sigma n_{BN})^2}e^{i4\pi\omega n_{BN} d}}, \quad
    A = (1-r_1^2)e^{-4(\pi \omega\sigma(n_{BN}-1))^2} + r_1^2e^{-8(n_{BN}^2+1)(\pi\omega\sigma)^2} \label{eq:r}
\end{equation}
To consider the porosity of the material, we use a simple model in which we decrease the dielectric function with the parameter $a$.
\begin{equation}
    \varepsilon_{BN}' = 1 + a(\varepsilon_{BN} - 1), \quad a \in (0,1) \label{eq:a}
\end{equation}
When Eq.~\ref{eq:r}-\ref{eq:a} are introduced it is possible to model a curve which describes the experimental data much better as presented in FIG.~\ref{fig:FTIR_limits}b. The fitting procedure revealed the thickness of the hBN layer $d_{BN}= 1122$~nm, surface roughness $\sigma = 46$~nm, porosity $a = 0.25$. The simulation presented in FIG.~\ref{fig:FTIR_limits}b inset suggests that using our method it would be possible to analyze crystals that are tens of micrometers thick. All this shows high universality of our method, which allows analyzing layers in the range from single atomic layers up to several $\mu$m.

\begin{figure}[htbp]
\includegraphics[width=\textwidth]{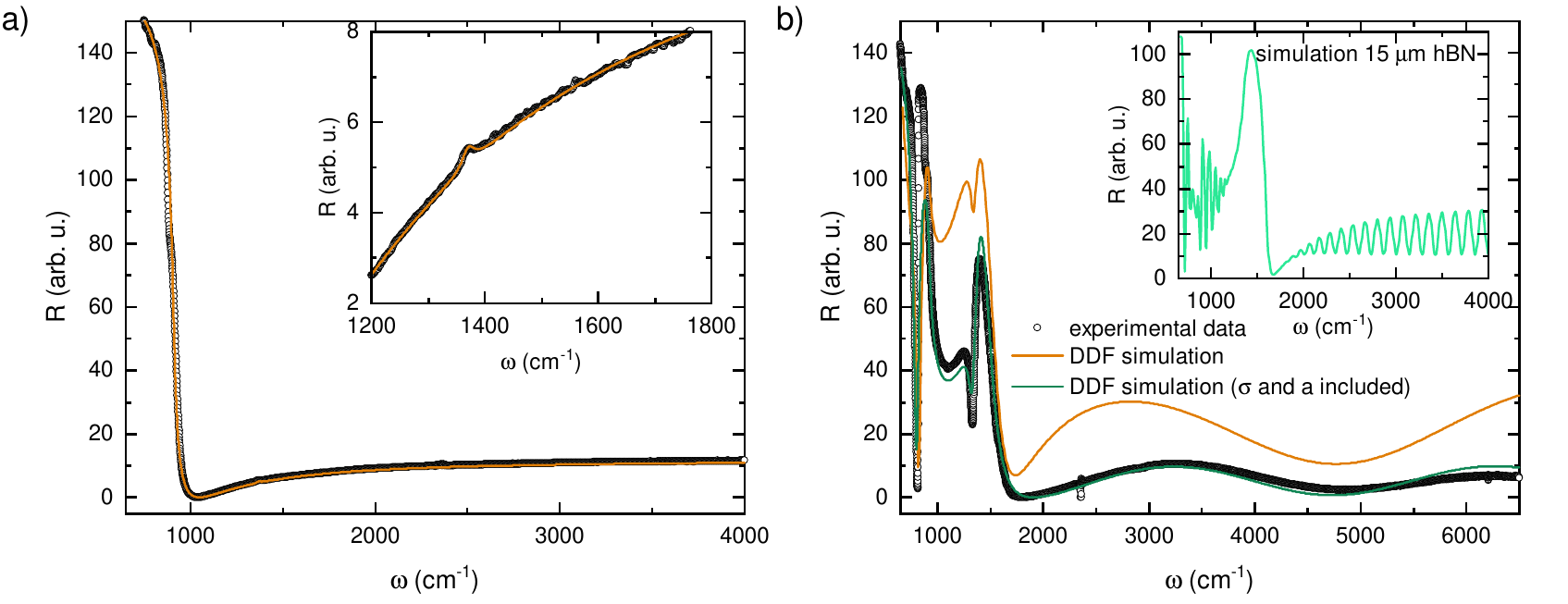}
\caption{\label{fig:FTIR_limits} FTIR spectra of hBN a) monolayer (zoom on hBN $E_{1u}$ vibrational mode peak), (b) 1122~nm thick rough and porous layer on Al$_2$O$_3$ substrate. Lines represent DDF simulations with and without including roughness and porosity. Inset shows simulation of 15~$\mu$m thick good quality hBN.}
\end{figure}

\newpage
\section*{Simulation of temperature dependence of $D_2$ defect occupation}

We assume a three-level energy model ($\varepsilon_0 < \varepsilon_1 < \varepsilon_2$). The highest energy level is the conduction band, which is represented by its high degeneracy $g_2 \gg 1$. The middle energy level can be degenerated as well ($g_1 \sim 1$). Occupation is governed by the Boltzmann distribution. We allow only thermal electron excitation. Then the occupation $n_i$ for the following energy levels will be given by:
\begin{gather*}
n_0 = N \frac{1}{Z}, \quad n_1 = N \frac{g_1e^{-e_{01}/kT}}{Z}, \quad N\frac{g_2e^{-e_{02}/kT}}{Z} \\
Z = 1 + g_1e^{-e_{01}/kT} + g_2e^{-e_{02}/kT}
\end{gather*}
where $T$-temperature, $k$-Boltzmann constant, $e_{ij}$-energy distance between states $i$ and $j$. For the case of analyzed defects in hBN: $e_{02} < 1$~eV, $e_{01} < e_{12}$ \cite{Weston2018, Vokhmintsev2019}. FIG.~\ref{fig:occupation} presents the results of simulations with different sets of modeled parameters. The values of the parameters for each simulation are presented in TAB.~\ref{tab:par_sim}. One can observe the simulated peak movement and its intensity change while changing parameters. The curve and parameters related to Sim$_0$ correspond to the simulation presented in the main text in FIG.~4. 

\begin{figure}[!htbp]
\includegraphics[]{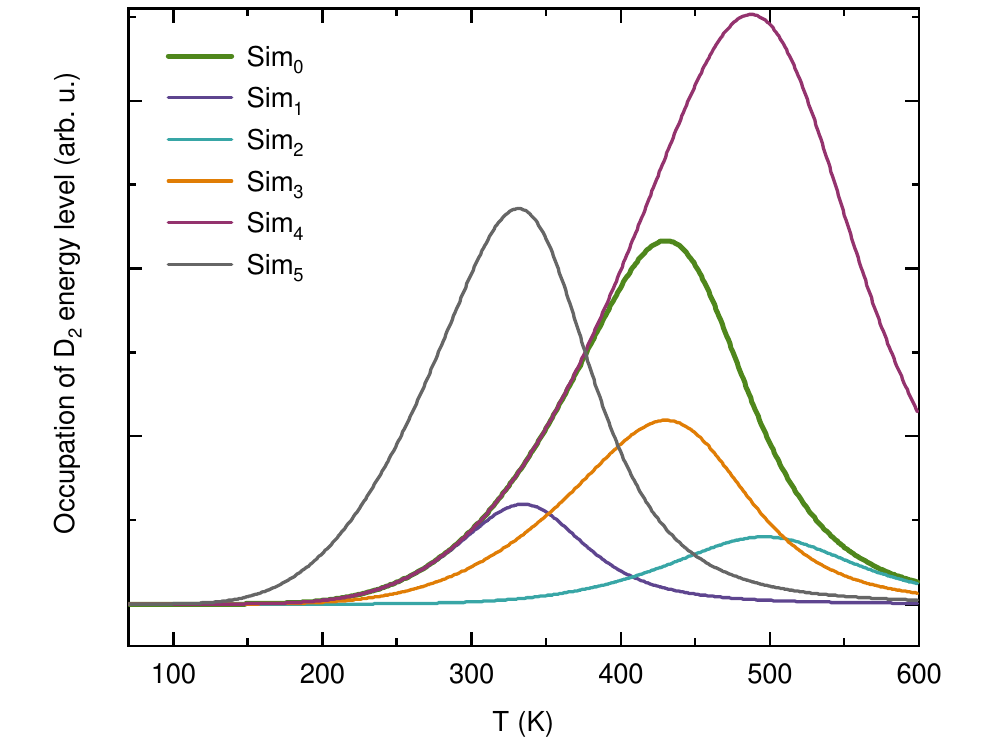}
\caption{\label{fig:occupation} Simulation of occupation of the middle energy level which we named $D_2$ defect. The values of modeled parameters are presented in TAB.~\ref{tab:par_sim}.}
\end{figure}

\begin{table*}[htbp]
\caption{\label{tab:par_sim}
Values of parameters for the curves simulated in FIG.~\ref{fig:occupation}.}
%\begin{ruledtabular}
\begin{tabular}{lcccc}
\hline\hline
Simulation &$e_{01}$ (eV)&$e_{12}$ (eV)&$g_1$ & $g_2$\\
\hline
Sim$_0$ & 0.158 & 0.573 & 2 & $10^8$  \\
Sim$_1$ & 0.158 & 0.400 & 2 & $10^8$  \\
Sim$_2$ & 0.250 & 0.573 & 2 & $10^8$  \\
Sim$_3$ & 0.158 & 0.573 & 1 & $10^8$  \\
Sim$_4$ & 0.158 & 0.573 & 2 & $10^7$  \\
Sim$_5$ & 0.100 & 0.400 & 1 & $10^7$  \\
\hline\hline
\end{tabular}
%\end{ruledtabular}
\end{table*}

\newpage

\section*{Anomaly observed for different samples}

\begin{figure}[htbp]
{\includegraphics{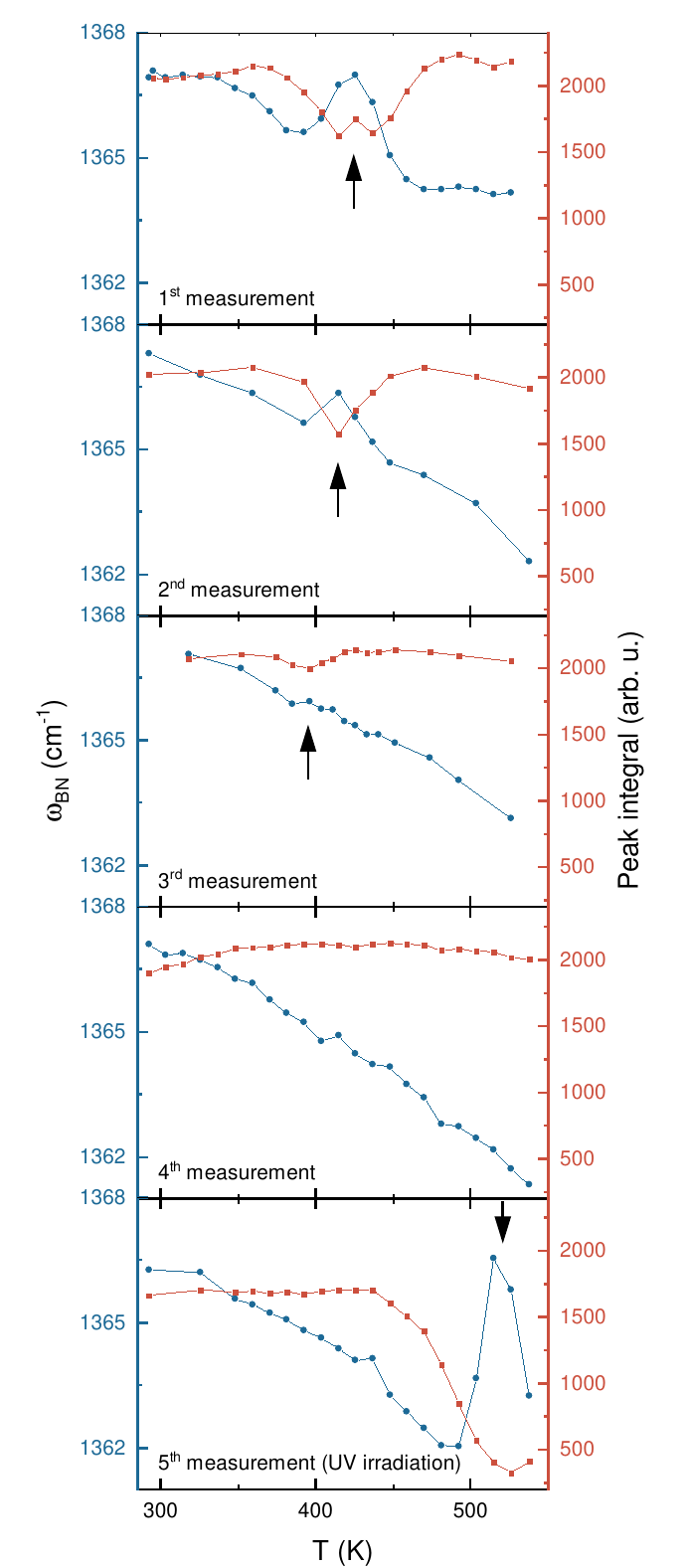}}
 \caption{\label{fig:anomaly_peak-integral} $E_{1u}$ phonon energy (blue circles) and peak integral coming from Lorentzian fitting (orange squares) as a function of temperature for the 20~nm thick sample grown in the Two-stage Epitaxy regime.  Arrows point out eh anomalous behaviour occurrence. Initial 4 temperature-dependent measurements lead to softening of the anomalous phonon energy increase. After UV irradiation the anomalous effect is enhanced. Lines are guide to the eye only. Due to the strong correlation between peak integral and other fitted parameters, there is no direct relation between peak integral intensity and the harmonic oscillator strength. At the temperatures at which the anomaly occurs, a peak integral dip is observed. For the measurement after UV irradiation, the minimal peak integral value is 5 times lower than the average for this temperature run.}
\end{figure}

\begin{figure}[!htbp]
\includegraphics[width=\textwidth]{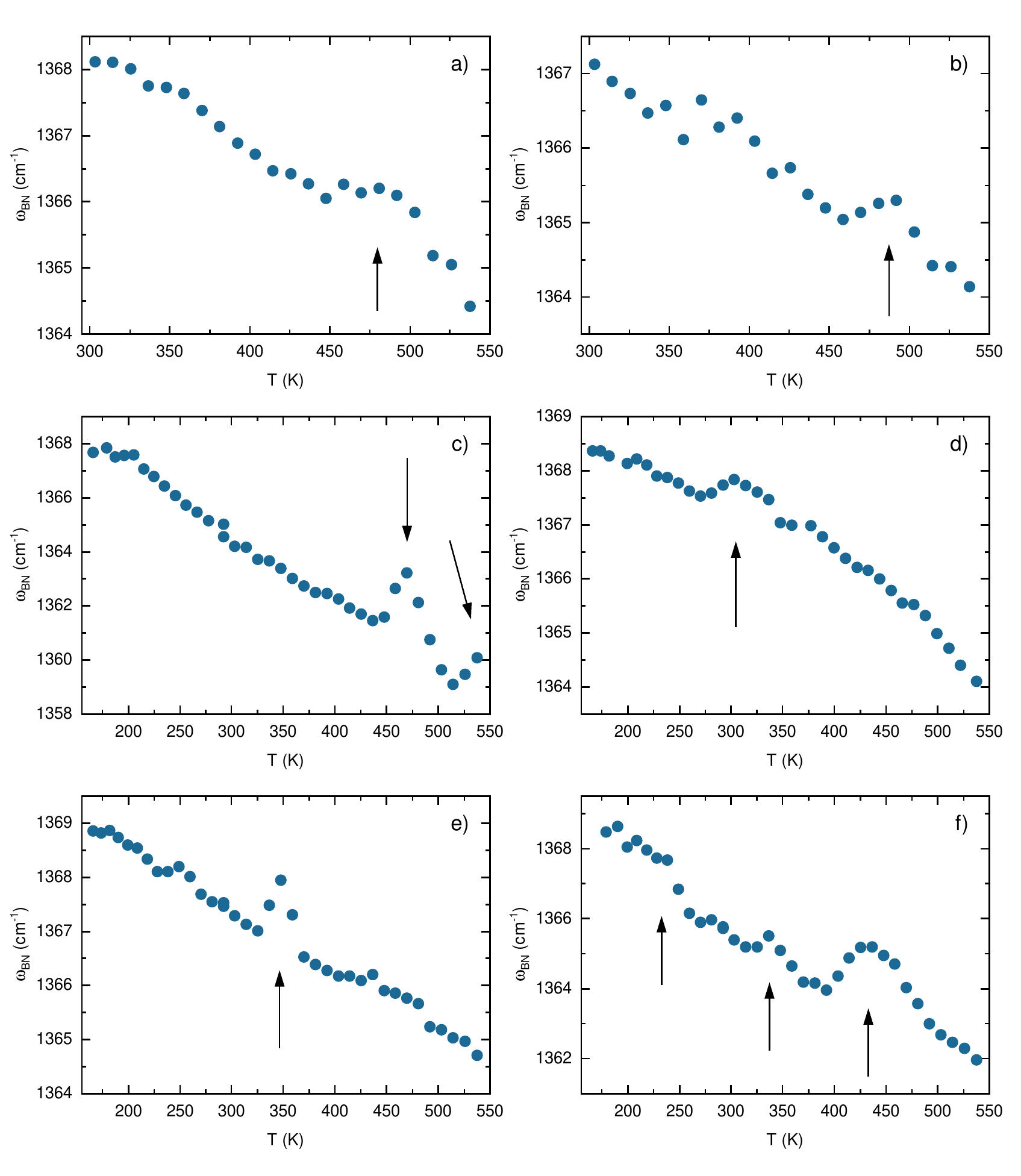}
\caption{\label{fig:anomaly-suppl} Anomalous behaviour of $E_{1u}$ phonon energy as a function of temperature for the samples grown in different conditions. Growth parameter details for each sample are presented in TAB.~\ref{tab:par_growth}. Arrows point out the anomalous behaviour occurrence.}
\end{figure}

\begin{table*}[htbp]
\caption{\label{tab:par_growth}
Growth parameters for the samples presented in FIG.~\ref{fig:anomaly-suppl}. The subscripts refer to the corresponding panels in FIG.~\ref{fig:anomaly-suppl}. Sample $S_c$ comes from the same growth process as the 20~nm thick Two-stage Epitaxy sample from the main text, but before measurement it was delaminated. The same for the pair of samples $S_e$ (as-grown) and $S_f$ (delaminated). The abbreviations used in the table for growth modes come from: CFG - Continous Flow Growth, FME - Flow Modulation Epitaxy, Two-stage - Two-stage Epitaxy. $T_g$ is the growth temperature.}
%\begin{ruledtabular}
\begin{tabular}{lccccc}
\hline\hline
Sample&Growth mode&$T_g$ ($^\circ$C)&NH$_3$ flow (ccm)&TEB flow (ccm)& $d_{BN}$ (nm)\\
\hline
$S_a$ & FME & 1295 & 4000 & 15 & 10.7 \\
$S_b$ & CFG & 1315 & 4000 & 10 & 5.2 \\
$S_c$ & Two-stage & 1309/1300 & 4000 & 10/15 & 15.4 \\
$S_d$ & Two-stage & 1305/1300 & 4000 & 10/15 & 36.8 \\
$S_e$ & Two-stage & 1295/1285 & 4000 & 10/15 & 8.5 \\
$S_f$ & Two-stage & 1295/1285 & 4000 & 10/15 & 8.0 \\
\hline\hline
\end{tabular}
%\end{ruledtabular}
\end{table*}

\begin{figure}[htbp]
{\includegraphics{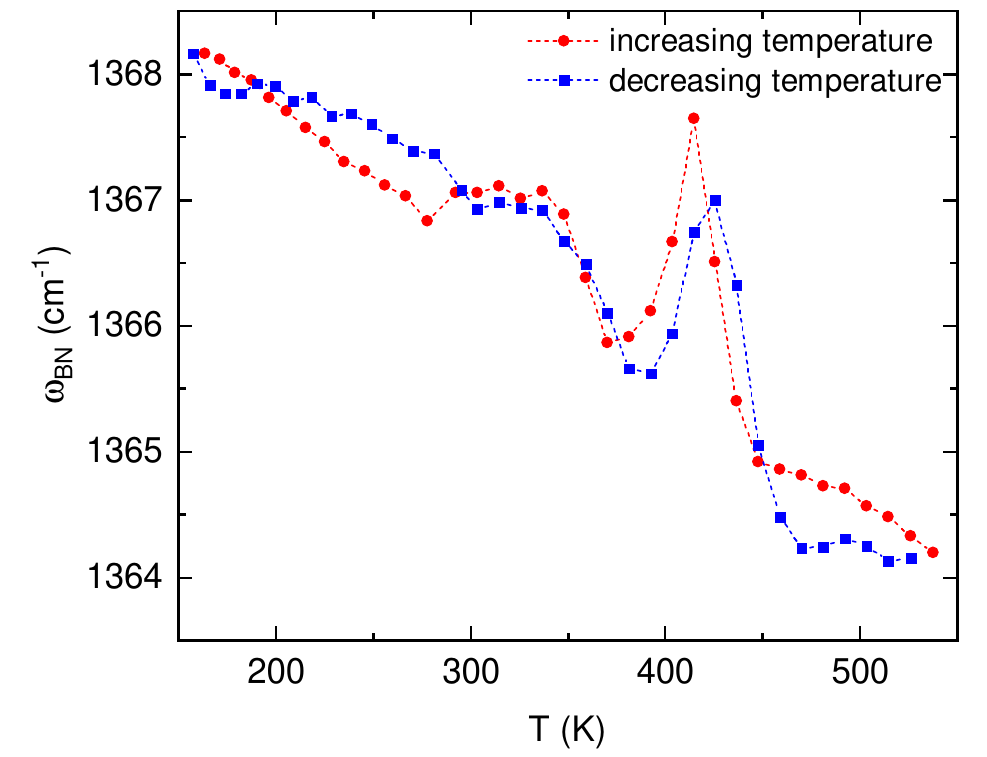}}
 \caption{\label{fig:anomaly_temperatures} $E_{1u}$ phonon energy as a function of temperature measured for increasing (red circles) and decreasing temperature (blue squares). Results refer to 1$^{\mathrm{st}}$ measurement from the main text. Dashed lines are guide for eye only.}
\end{figure}

\clearpage

\end{widetext}

\end{document}